\documentclass[final,5p,times]{elsarticle}

\usepackage{amssymb}
\usepackage{graphicx}
\usepackage{amsmath}
\usepackage{xcolor}
\usepackage{subfig}
\usepackage{array,multirow}

\journal{Computer Science}

\begin{document}

\begin{frontmatter}



\title{A fast and practical grid based algorithm for point-feature label placement problem }


\author[yoa]{Yasemin \"{O}zkan Ayd{\i}n \corref{cor1}}
\ead{yesminozkan@yahoo.com}
\author[kl]{Kemal Leblebicio\u{g}lu }

\cortext[cor1]{Corresponding author}
\address[yoa]{School of Physics,  Georgia Institute of Technology, Atlanta, USA}
\address[kl]{Department of Electrical and Electronics Engineering, Middle East Technical University, Ankara, Turkey}


\begin{abstract}
Point-feature label placement (\textit{PFLP}) is a major area of interest within the filed of automated cartography, geographic information systems (\textit{GIS}), and computer graphics. The objective of a label placement problem is to assign a label to each point feature so as to avoid conflicts, considering the cartographic conventions. According to computational complexity analysis, the labeling problem has been shown to be NP-Hard. It is also very  challenging to find a computationally efficient algorithm that is intended to be used for both static and dynamic map labeling. In this paper, we propose a heuristic method that first fills the free space of the map with rectangular shape labels like a grid and then matches the corresponding point feature with the nearest label. The performance of the proposed algorithm was evaluated through empirical tests with different dataset sizes. The results show that our algorithm based on grid placement of labels is a useful, fast and practical solution for automated map labeling.   

\end{abstract}


\end{frontmatter}

\newcommand{\rfig}[1]{Fig.~\ref{#1}}
\newcommand{\rtab}[1]{Table~\ref{#1}}
\newcommand{\rsec}[1]{Sec.~\ref{#1}}
\section{Introduction}
\label{intro}

Visualization of information on graphical displays  is a very important task when producing user-friendly, informative maps. Labels are an essential part of the maps when identifying point (e.g., cities, towns, mountains), line (e.g., streets, rivers), or area (e.g., countries, oceans) features. Point-feature label placement (\textit{PFLP}) is a challenging problem in the area of automated cartography and geographic information systems (GIS). The aim is to place labels with a certain shape near to corresponding point features while considering cartographic rules such as \cite{Imhof.1975,Yoeli.1972};
\begin{itemize}
	\item The size of the labels must suitable the text written in it,
	\item No overlaps with other labels or features,
	\item The connection between label and its associated feature should be clear,
	\item The algorithm should be fast and accurate,
	\item A label must be placed in the best possible location.
\end{itemize}
  
Although humans are successful in overcoming the basic labeling problems such as conflict and uncertainty, obtaining a map or drawing which has labels perfectly placed on it is very time consuming and non-trivial to do manually. Therefore, developing computer algorithms for automated label placement has received much attention by scientists in a wide range of fields, particularly cartography, architecture, computational geometry,  image analysis, and navigation systems.\\

To display information about objects in the interactive map such as type of aircraft, name of buildings in a dangerous area, the type of military supplies that aircraft carry or to draw attention to a hazardous area, labeling process must be done quickly and automatically. Especially, in real-time applications where users can change the scale and viewpoint of the map continuously, run-time of the algorithm is a very critical factor that should be considered. In most algorithms a considerable amount of time is spent in detecting label-label or label-feature overlaps \cite{Kakoulis.2006}. If too many objects are close in a screen, the labels causes cluttering or some objects are not labeled properly. Rather than produce results that obey all good labeling steps, our goal is to guarantee that all objects are labeled adequately in the  map. The major limitation of the present study is that all labels should have common size and type.\\

Simulated annealing (\textit{SA}) is the most commonly used cartographic labeling algorithm. It is an energy based iterative and stochastic global searching algorithm \cite{Zoraster.1997,Kirkpatrick.83}. Genetic algorithm (\textit{GA}) has been applied to solve various optimal problems. It has been shown that \textit{SA} and \textit{GA} exhibit the best performance in terms of non-conflict labeling point ratio, but \textit{SA} produces a faster solution than \textit{GA} when the node number is increased \cite{Hong.2004,Raidl98agenetic,Dijk.2002}.\\

Generally, the cartographic labeling algorithm consists of three subtasks; (1) label candidate position selection, (2) cost evaluation, (3) label assignment \cite{Edmondson.1996}. The candidate label that touch the point feature can be placed at an 1, 2, 4 or 8 fixed position or moved continuously around the node. After all candidate label positions are defined, the conflict graph is obtained based on overlaps between the labels and nodes. The optimization algorithms or heuristic methods then find the best label configuration with a minimum overlap considering cartographic preferences. If the algorithm can not obtain a result without conflict, some labels can be removed. The time required to select candidate label positions specifies the quality (computation time) of the algorithm. Our algorithm first fills the free space of map with an evenly-spaced axis parallel rectangular labels. This gives a conflict-free candidate label set (\textit{CLS}) and if the number of label in \textit{CLS} is greater than or equal to node number then all nodes can be labeled without any label-label or label-node overlap. The main contribution of our paper is to solve the conflict problem at the phase of selection of candidate label set where other algorithms in the literature solve it after obtaining a candidate label set.\\

If the density of the points on the map to be labeled does not allow to place labels without conflict, a leader line may be used to show the correspondence between point and label \cite{Zoraster.1997,Vollick.2007,Bekos.2005}. In this case, labels are placed away from the point and a straight line or a combination of parallel and orthogonal lines connects point to label \cite{Bekos.2005,Bekos.2006,Bekos.2007}. The objective is to find a minimum length leader without overlap \cite{Bekos.2007,Bekos.2008,Bekos.2009,Kindermann.2015}. The length of the leader is important, since the shorter the leader line the smaller is the probability of two lines intersecting \cite{Wang.2013}. The ports where leaders touch labels may be prescribed or may be arbitrary \cite{}. Most of the studies draw a frame around the map, and place the labels outside of this frame by either one \cite{Bekos.2005,Bekos.2006,Benkert.2008}, two or four side \cite{Bekos.2007,Bekos.2009,Kindermann.2015}, whereas our study allows the placement of leader-connected labels not only at the boundary but anywhere in the map where there is empty space.\\  

In this paper, we propose an efficient and simple heuristic
method that can be also used in real time applications and report
on a series of empirical tests to show its performance. The aim
is to obtain the best label positions in a predefined map without
any overlap. The input of the system is n point features and
corresponding labels whose size are known. The outputs are
placement of labels in a map and connection of labels with associated features with the shortest line. This simple algorithm can be used in the field of cartography, computational geometry, or information visualization. \\

\section{Grid Based Label Placement Algorithm}
\label{sec:GBLP_algorithm}

\begin{figure}[h]
	\centering
	\includegraphics[width=1\linewidth]{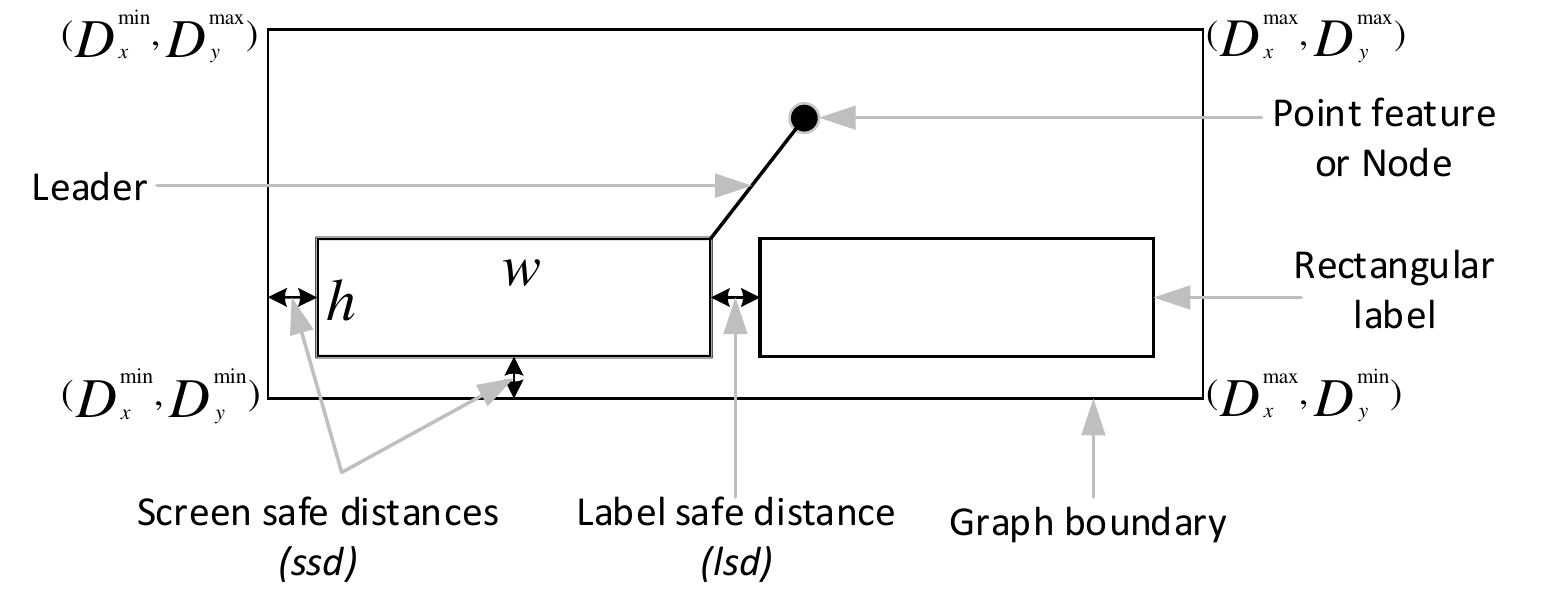}
	\caption{The graphical illustration of the graph boundary and definitions used in the paper. A leader line is used to create a visual connection between the label and its corresponding point feature. We left some space between the graph boundary and labels to reduce the ambiguity.}
	\label{fig:fig_definitions}
\end{figure}

\begin{figure}[h]
	\centering
	\includegraphics[width=0.55\linewidth]{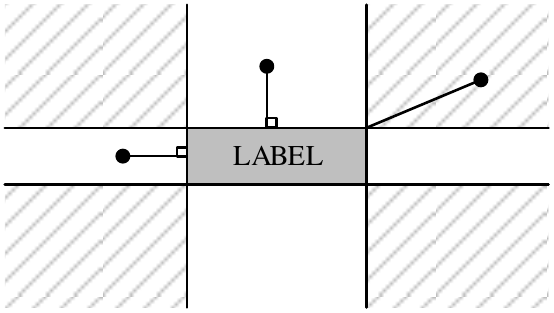}
	\caption{The leader is connected to its corresponding label with the shortest line. If a point feature is in the filled area, a leader connects the point feature to the nearest corner of its corresponding label.}
	\label{fig:fig_leader}
\end{figure}

In this section, we introduce the terminology used throughout  the paper and explain the details of grid based algorithm intended to be usedfor labeling of point features. The graphical illustration of the labeling problem is given in \rfig{fig:fig_definitions}. A leader line (see \rfig{fig:fig_leader}) is used to show the correspondence between the label and point feature. The graph boundaries are defined as $(D_x^{min},D_x^{max})$ and $(D_y^{min},D_y^{min})$. We leave some distance between labels and graph boundary to clearly identify labels from the edges of graph.  Other terms used throughout the paper are given in \rtab{tab:terms}.\\

The input of the labeling problem consists of a set $ P=\left\{p_1,p_2,...,p_n \right\}\subseteq \mathbb{R}^2 $ of $n$ randomly generated point features where $p_i =(p_{i_x},p_{i_y}), i=1,2,...,n$. Each point $p_i$ is associated with an axis-parallel rectangular label $l_i$ of width $w$ and height $h$. $\mathcal{L}$ is the set of all label positions, $\mathcal{L}_p$ is the set of top-k closest labels of all point features to be labeled, $\mathcal{L}_{p_i}$ is the set of top-k closest labels of point feature $p_i$ of $P$.  The task is to assign a label to each point feature in 2-dimensional space from the set $\mathcal{L}$. A label should be close to the point to which it belongs, and should not overlap with other labels and graphical features. Additionally, the center of each label $(l_{i_x},l_{i_y})$ in the set $\mathcal{L}$ must satisfy the constraints of the graph boundaries,

\begin{eqnarray}
D_x^{min}+\frac{w}{2} <l_{i_x}<D_x^{max}-\frac{w}{2} \\ \nonumber
D_y^{min}+\frac{h}{2} <l_{i_y}<D_y^{max}-\frac{h}{2}.
\end{eqnarray}



 Different from the algorithms \cite{Christ.1995,Yamamoto.02} that place a finite number of positions being tangential to the point feature or slider model \cite{vanKreveld.1999} that allow any position on the edges of label, our method is based on placing as many axis-parallel rectangular labels of fixed height and width as possible in a predefined  map without overlapping. The label placement is similar to asymmetric graph paper which has some space within each division. The labeling process can be subdivided into three stages:
 \begin{enumerate}
 	\item Calculation of potential label positions,
 	\item Ranking of the labels according to their distances to graphical features,
 	\item Assignment of labels to the corresponding point features.
 \end{enumerate}
\begin{table}[t]
	\scriptsize
	\centering
	\caption{The meaning of the terms used in the paper.	 }
	\begin{tabular}{|p{4cm}|p{4cm}|}
		\hline \rule[-2ex]{0pt}{5.5ex} \textbf{Point feature or Node} & A graphical feature to be labeled  \\ 
		\hline \rule[-2ex]{0pt}{5.5ex} \textbf{Leader} & The shortest line that connects a label to the corresponding point feature \\ 
		\hline \rule[-2ex]{0pt}{5.5ex} \textbf{Label closeness level} & The closeness order of the nearest n\textit{th} label to the corresponding point \\ 
		\hline \rule[-1ex]{0pt}{4.5ex} \textbf{Nearest Label Matrix (NLM)} & An n-by-k matrix that stores number of top-k closest labels of all point features \\ 
		\hline \rule[-2ex]{0pt}{5.5ex} \textbf{Label safe distance (LSD)} & A default horizontal and vertical distance between labels \\ 
		\hline \rule[-2ex]{0pt}{5.5ex} \textbf{Screen safe distance (SSD)} & A default distance between labels and screen \\ 		\hline 
	\end{tabular} 
	
	\label{tab:terms}
\end{table}

\begin{figure*}[!ht]
	\subfloat[]{%
		\fbox{\includegraphics[width=1\textwidth,width=16cm,height=9cm]{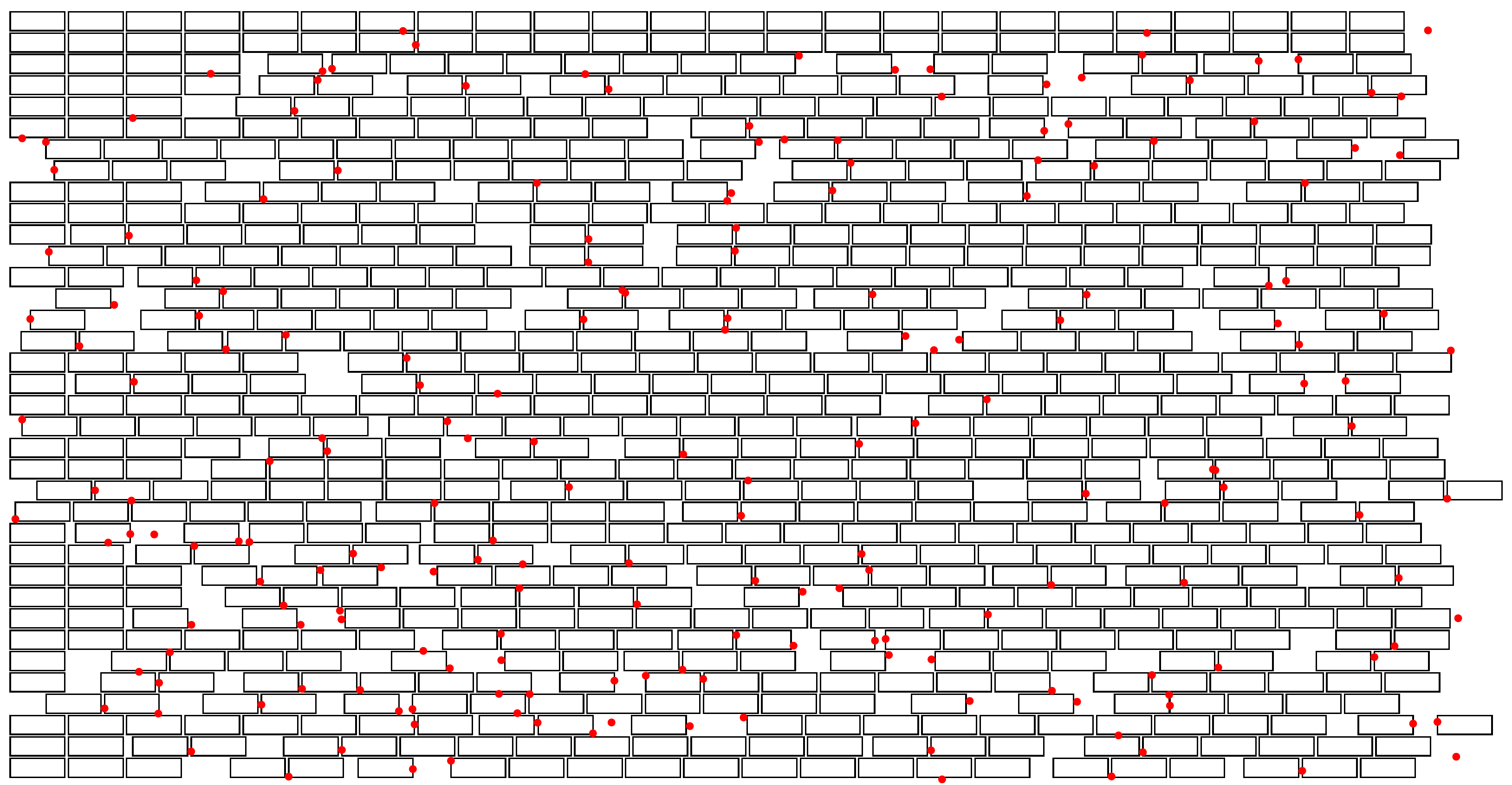}}
		\label{fig:screen_example}%
	}\\
	\subfloat[]{%
		\fbox{\includegraphics[width=1\textwidth,width=16cm,height=9cm]{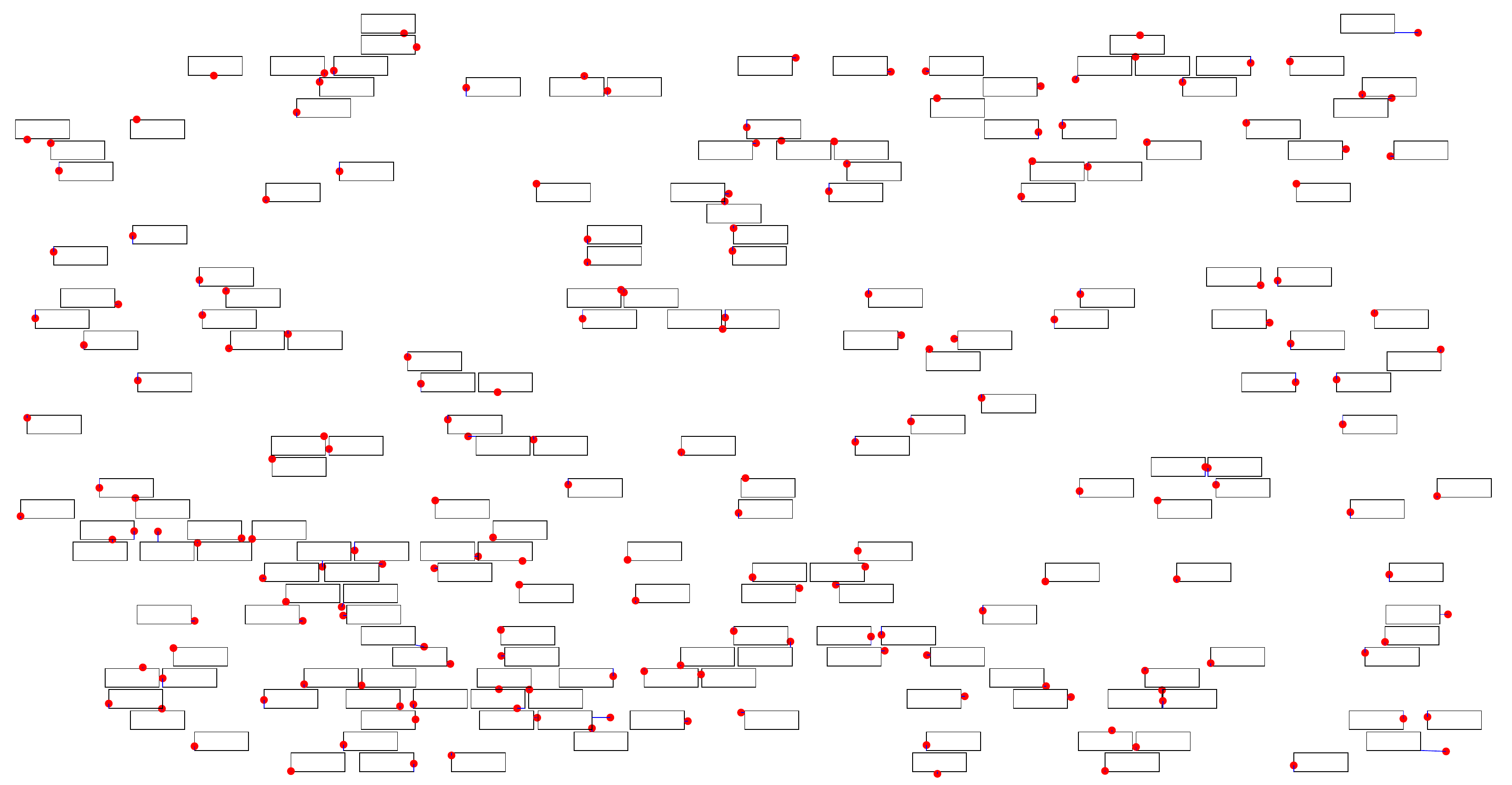}}
		\label{fig:label_assignment}%
	}
	\caption{(a) Potential label positions  (b) The label position after assignment}
\end{figure*}

\begin{figure}[th]
	\centering
	\includegraphics[width=1\textwidth,width=8cm,height=5cm]{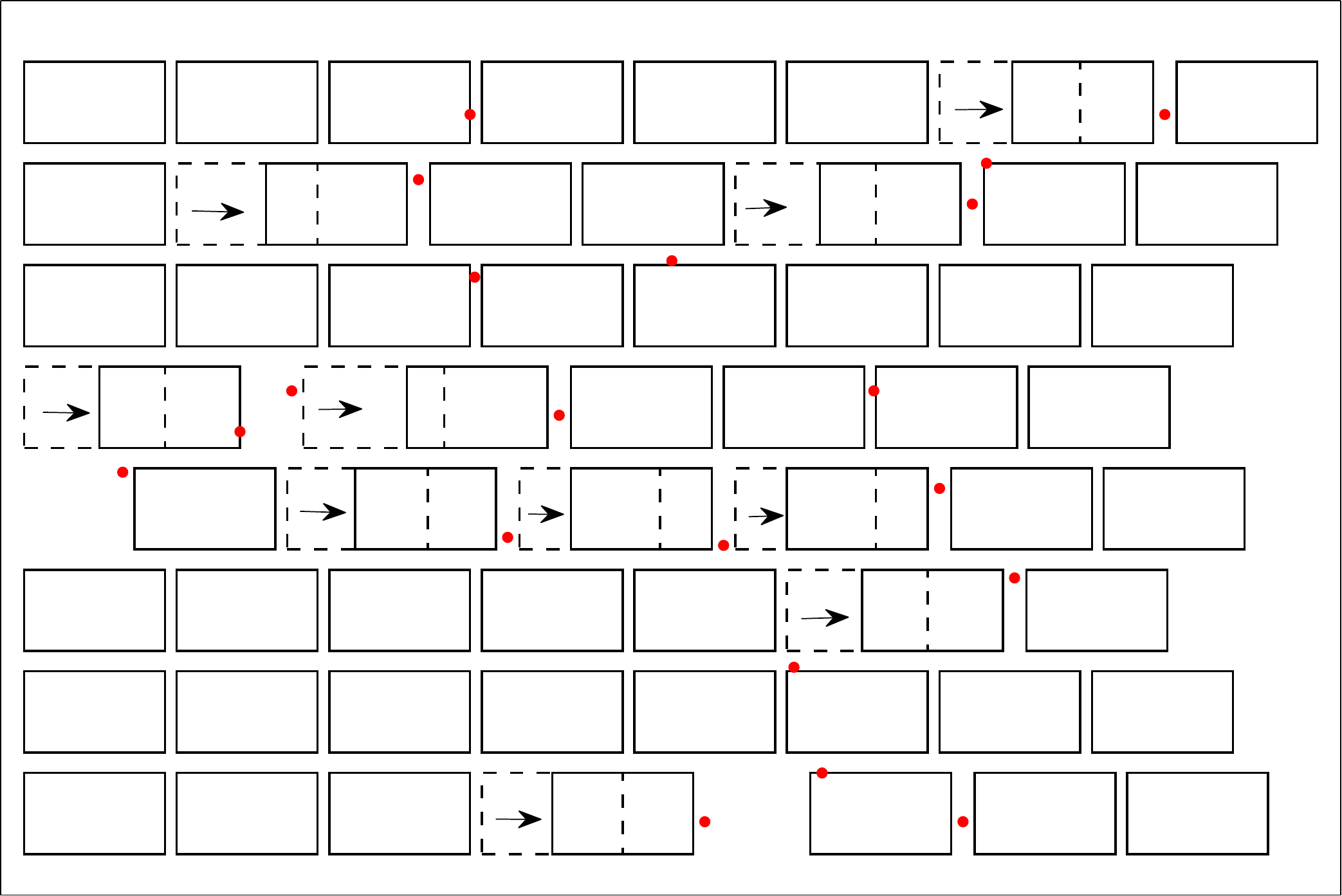}
	\caption{The sweep-phase of the algorithm. Labels are arranged side by side so that there is some space between them. The labels (dashed) located away from the node  are shifted along the x-axis from left to right to their final position (solid). 
		\label{fig:fig_arrow} }
\end{figure}
The detailed algorithm for the grid based label placement is as follows;
\begin{itemize}
\item Choose the set of randomly generated point features in the plane (In a real time application they are obtained from GPS data or user defined) and place them in a map.
\item Place fixed size rectangular type labels into the map side-by-side without overlap with other labels and nodes. The labels are arrayed in rows with some space between each other and positioned horizontally starting from the bottom left corner of the scene. In order to increase visibility and clearness, we leave horizontal and vertical white space between labels, called \textit{label safe distance}. If the horizontal distance, {\small $x_d$}, between the right edge of a label and the corresponding point feature is {\small $ LSD<x_d<LSD+w$}, then the label is shifted along the x-axis until {\small $x_d=LSD$ }without overlapping other map features which are in the vicinity of the point feature. We call this step as sweep-phase of the algorithm (see \rfig{fig:fig_arrow} ). 
\item Store the $(x,y)$ coordinates of four corners of all labels in an $m$x${\small 4}$ matrix, where $m$ is the label number.
\item Find the nearest corner of the each label for $n$ nodes by calculating the distance 
\begin{equation}
d_{ij} = \|x_i-x_j\|^2_2, ~ i=1,...,n ~ \text{and} ~ j=1,...,m
\end{equation}
between each node and the four corners of all labels.
\item Find the top-k nearest labels of each node and store label numbers and their position in the \textit{Nearest Label Matrix} (NLM).
\item First of all, the labels in the first row of NLM are assigned the nodes. At the end of the each assignment, we remove the number of assigned label from all rows and columns of the NLM. If we have unlabeled nodes after assignment of the first nearest labels, we continue with second nearest labels. This procedure continues until all nodes have one label. When a label closeness level is the same for more than one node, some leader lines can be overlap with other labels. In the final label assignment produced by our algortihm, each assigned label does not overlap any other label or node. 
\item After all nodes are connected with labels, the unused labels are erased and the rest are drawn on a screen with a leader line (see \rfig{fig:fig_leader}).
\end{itemize}
 \begin{figure*}[!ht]
 	\centering
 	\subfloat[]{%
 		\includegraphics[width=5.5cm]{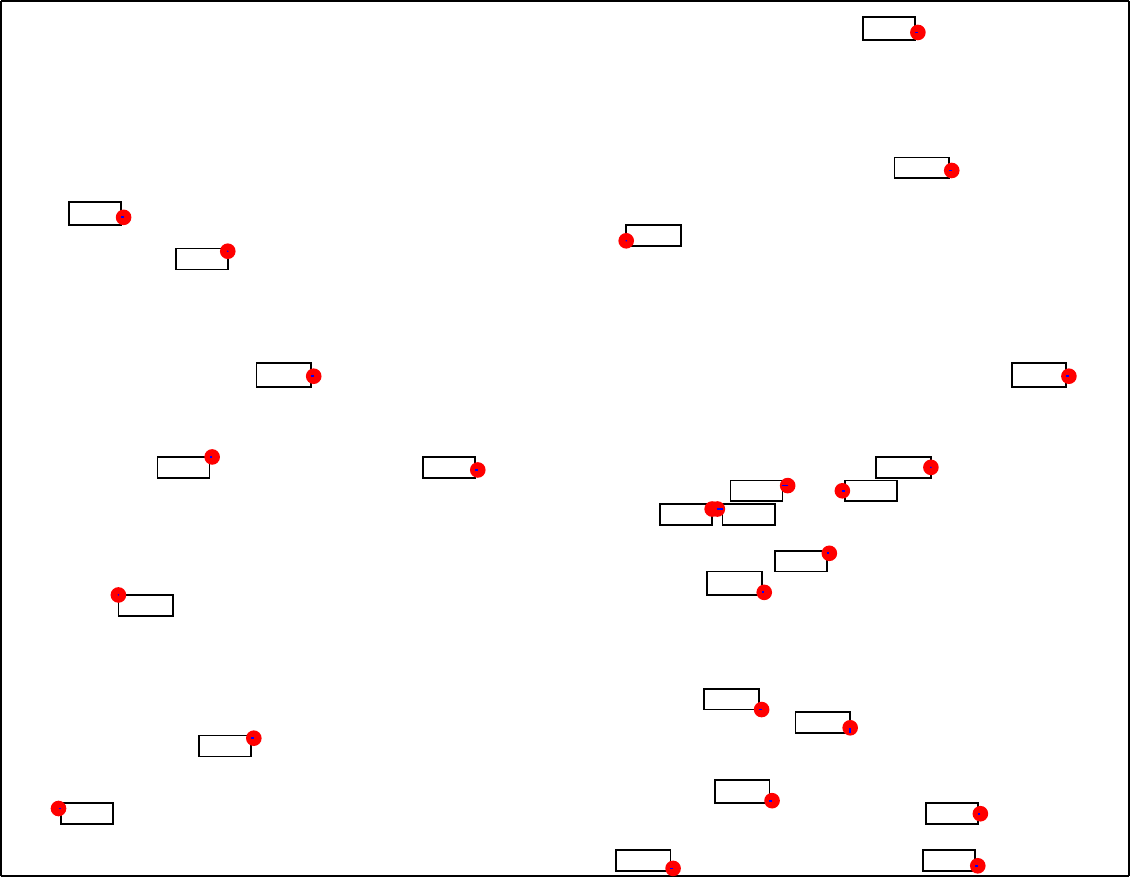}%
 		\label{fig:label_ass_25node}%
 	}~
 	\subfloat[]{%
 		\includegraphics[width=5.5cm]{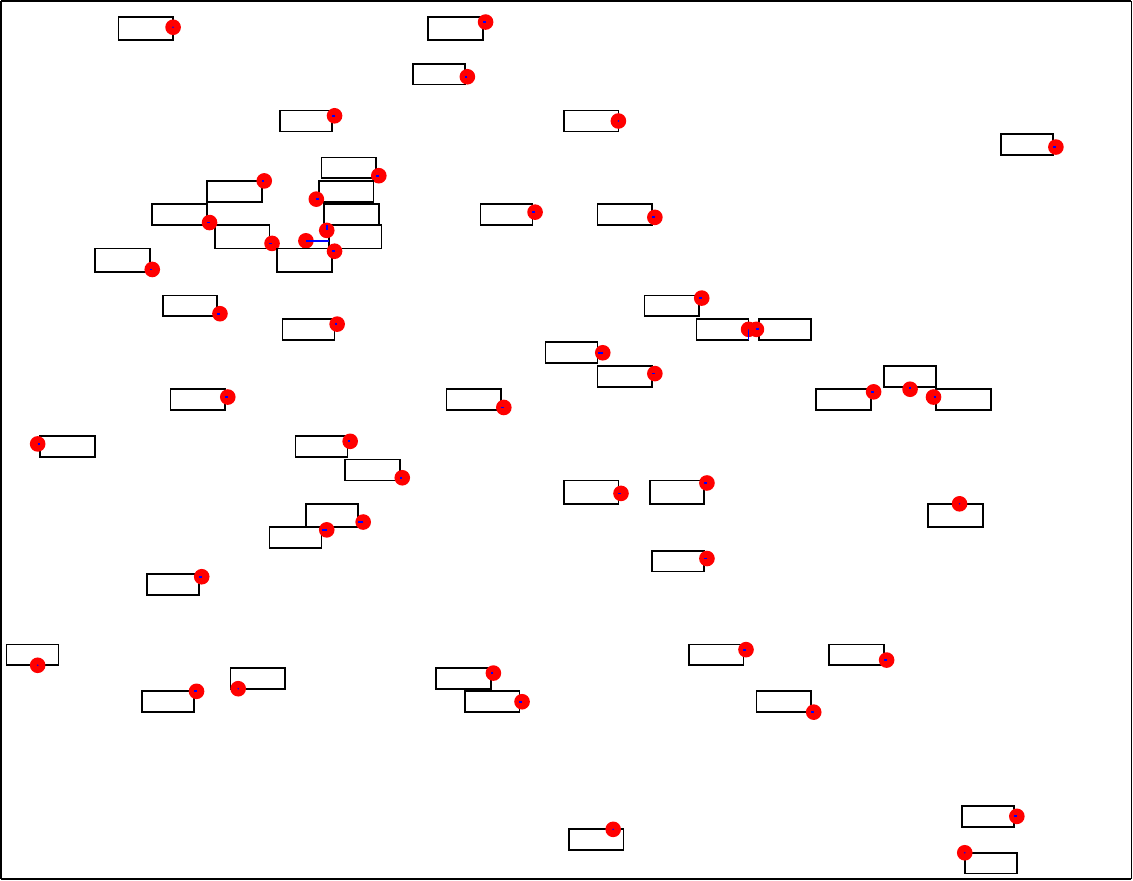}%
 		\label{fig:label_ass_50node}%
 	}~
 	\subfloat[]{%
 		\includegraphics[width=5.5cm]{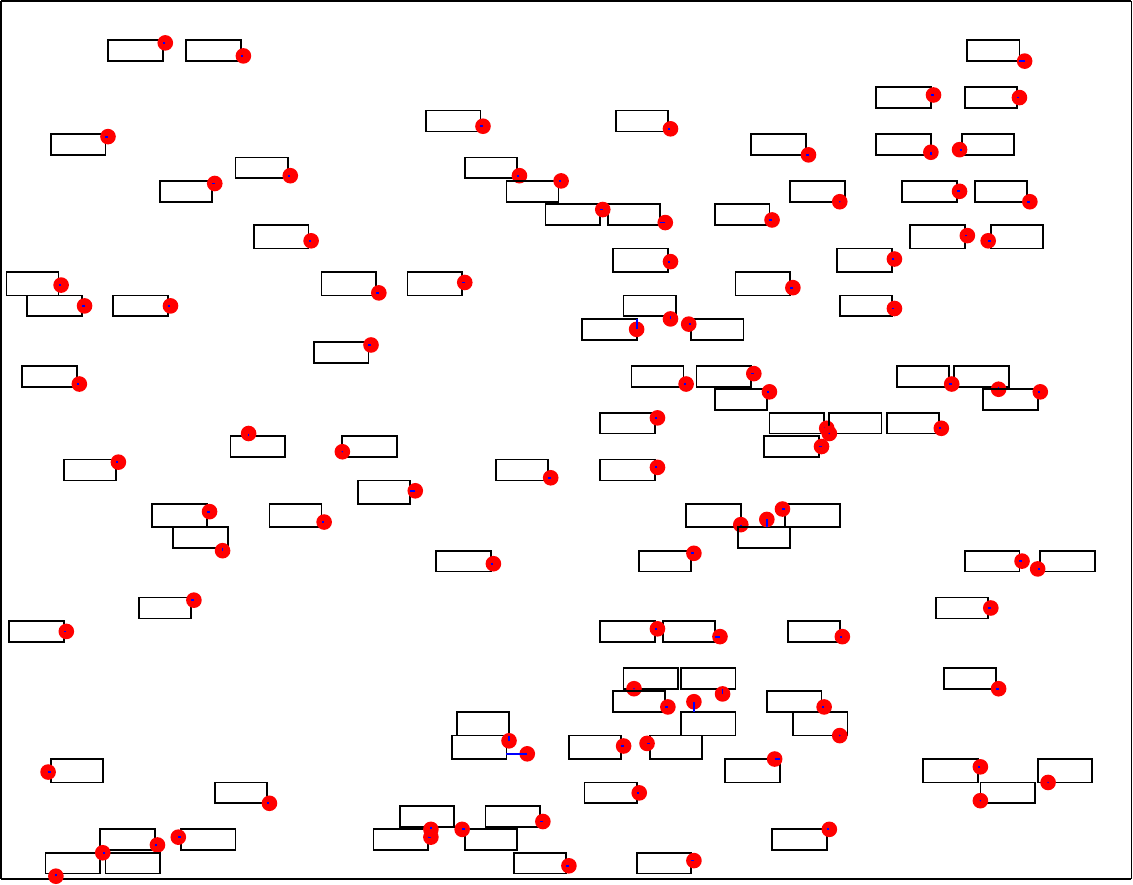}%
 		\label{fig:label_ass_100node}%
 	}\\
 	\subfloat[]{%
 		\includegraphics[width=5.5cm]{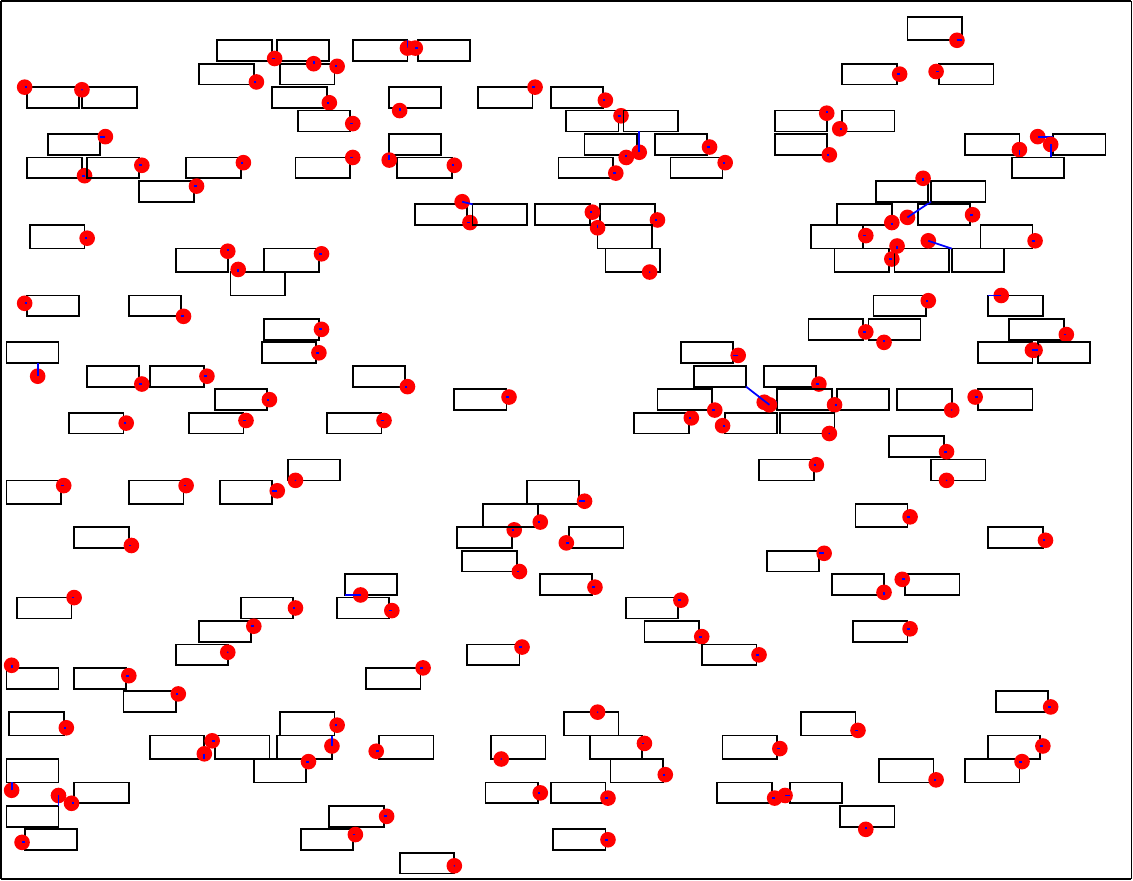}%
 		\label{fig:label_ass_150node}%
 	}~
 	\subfloat[]{%
 		\includegraphics[width=5.5cm]{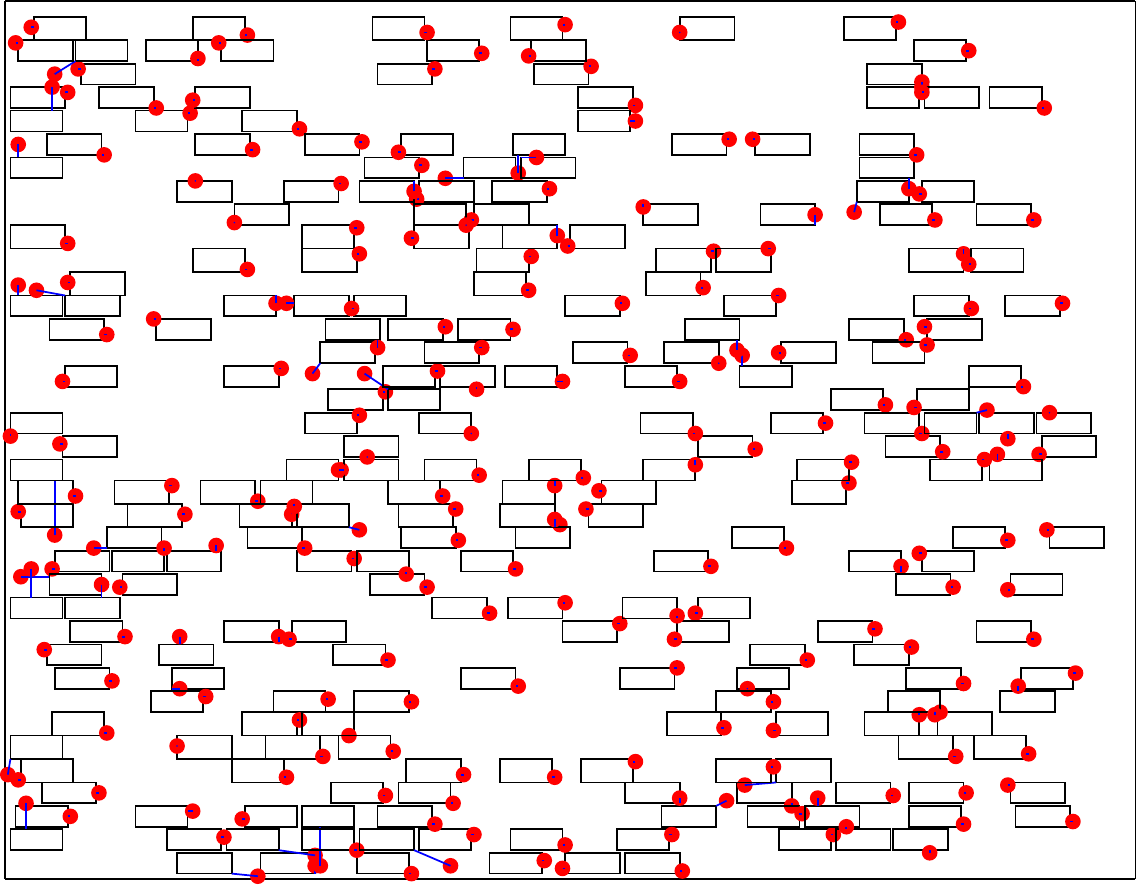}%
 		\label{fig:label_ass_250node}%
 	}~
 	\subfloat[]{%
 		\includegraphics[width=5.5cm]{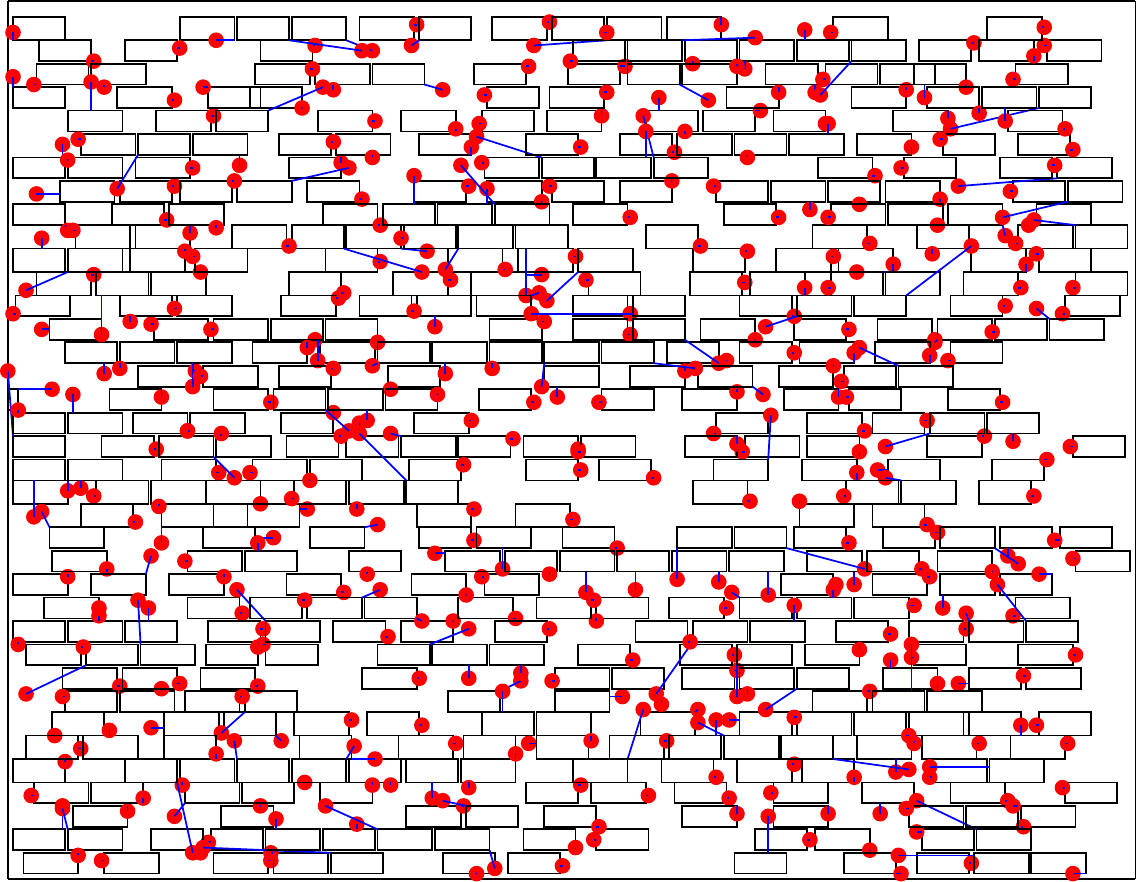}%
 		\label{fig:label_ass_500node}%
 	}
 	
 	\caption{The final label position of the randomly generated map with a node number (a)$ n=25 $, (b) $ n=50 $, (c) $ n=100 $, (d) $ n=150 $, (e) $ n=250 $, (f) $ n=500 $}
 	\label{fig:label_ass_allnode}
 \end{figure*}
\section{Results}

We have implemented the algorithm in MATLAB (R2011a) and all tests were run on an Intel(R) Core (TM) i7-2630 QM 2.00 Ghz CPU  with 4 Gb of RAM. We randomly placed $n$ nodes on a region of size $3000$ by $4000$. For the experiments, labels are axis-parallel rectangles and each graphical feature is associated with the same number of equal sized labels. In our implementation, the construction of an initial set $\mathcal{L}$ of label positions, the calculation of label-node distance, and the formation of the matching have been produced according to the method described in \rsec{sec:GBLP_algorithm}. According to node number, the label size can be adjusted to speed up label placement phase. We ran two sets of experiments
\begin{enumerate}[i]
	\item Label size is fixed, we changed the node number,
	\item Labels are rectangle, in successive runs of the algorithm we changed the height and width of the label.
\end{enumerate}

\begin{figure}
	\centering
	\includegraphics[width=0.8\linewidth]{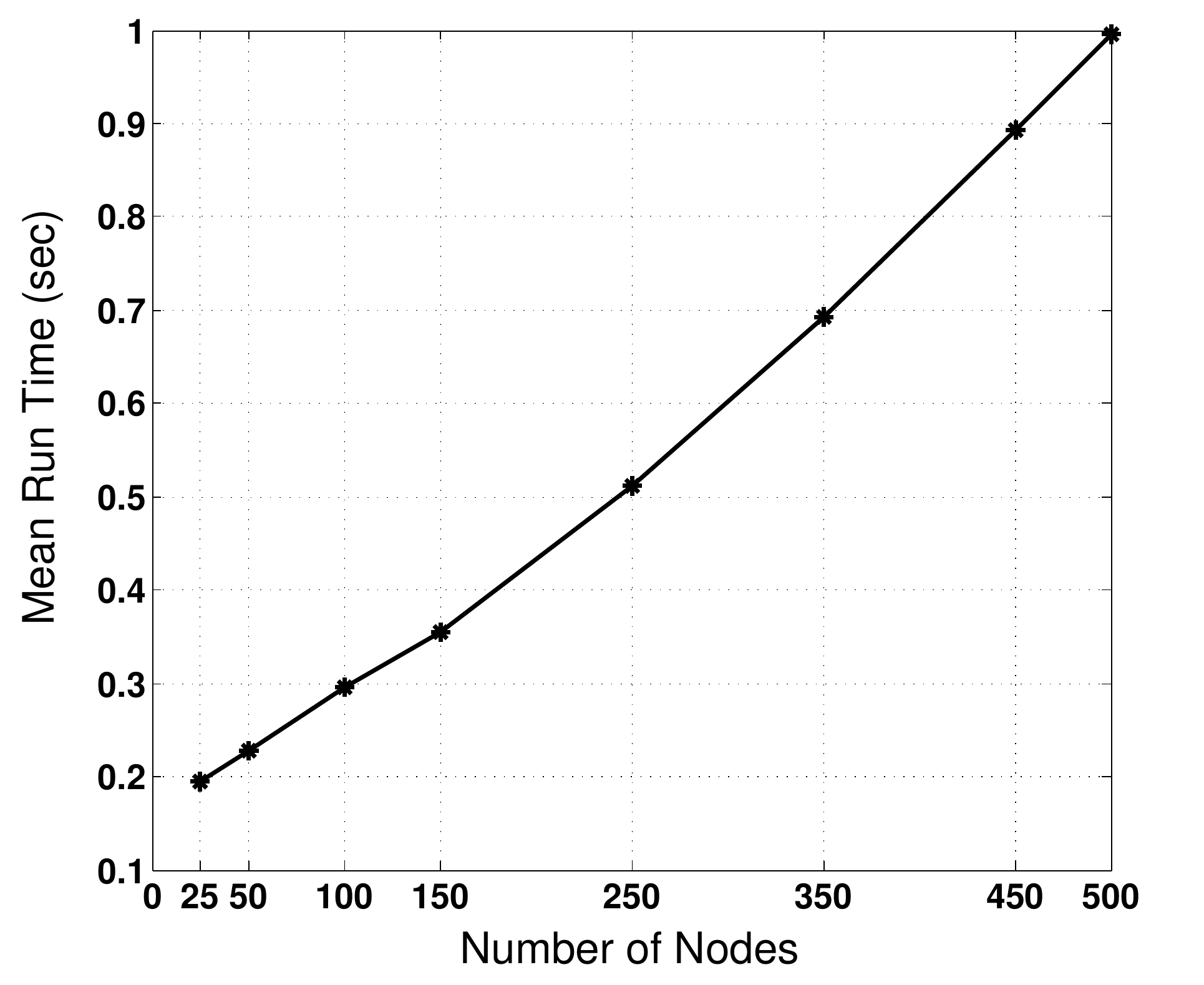}
	\caption{Results of empirical testing of the algorithm on randomly generated map data (label size is 150x100 units, map size is 3000x4000 units).The vertical axis shows the CPU time (in seconds) of algorithm for different node number. The results are averaged over a hundred trial.}
	\label{fig:LW150_LH100_SC3000X4000_meanruntime}
\end{figure}

In the first group of tests, we fixed the label size ($w=150,h=100$) and we looked the relation between node number and run time of the algorithm. To determine whether the performance of the algorithm was affected by the particular distribution of nodes, we conducted series of simulations with different node numbers. For each size of random datasets, we performed 100 trials, and the results were averaged. As seen in  \rfig{fig:LW150_LH100_SC3000X4000_meanruntime}, the label assignment procedure takes less than one minute for all datasets and there is an almost linear relationship between label number and run-time of the algorithm. The algorithm runs slower for smaller size labels because the initial set of label positions is much larger for smaller size labels. A huge amount of time is spent in filling the free space of map with labels at the beginning of the algortihm.  \rfig{fig:label_ass_allnode} shows screen-shot of the final label assignments for different node numbers.  Once the size of the labels increases above a certain threshold, the labeling quality decreases quickly since the label-leader line overlap increases. It will be an interesting problem to find efficient techniques that detect overlaps of labels with leader lines.\\

\begin{figure}[t]
	\centering
	\includegraphics[width=0.9\linewidth]{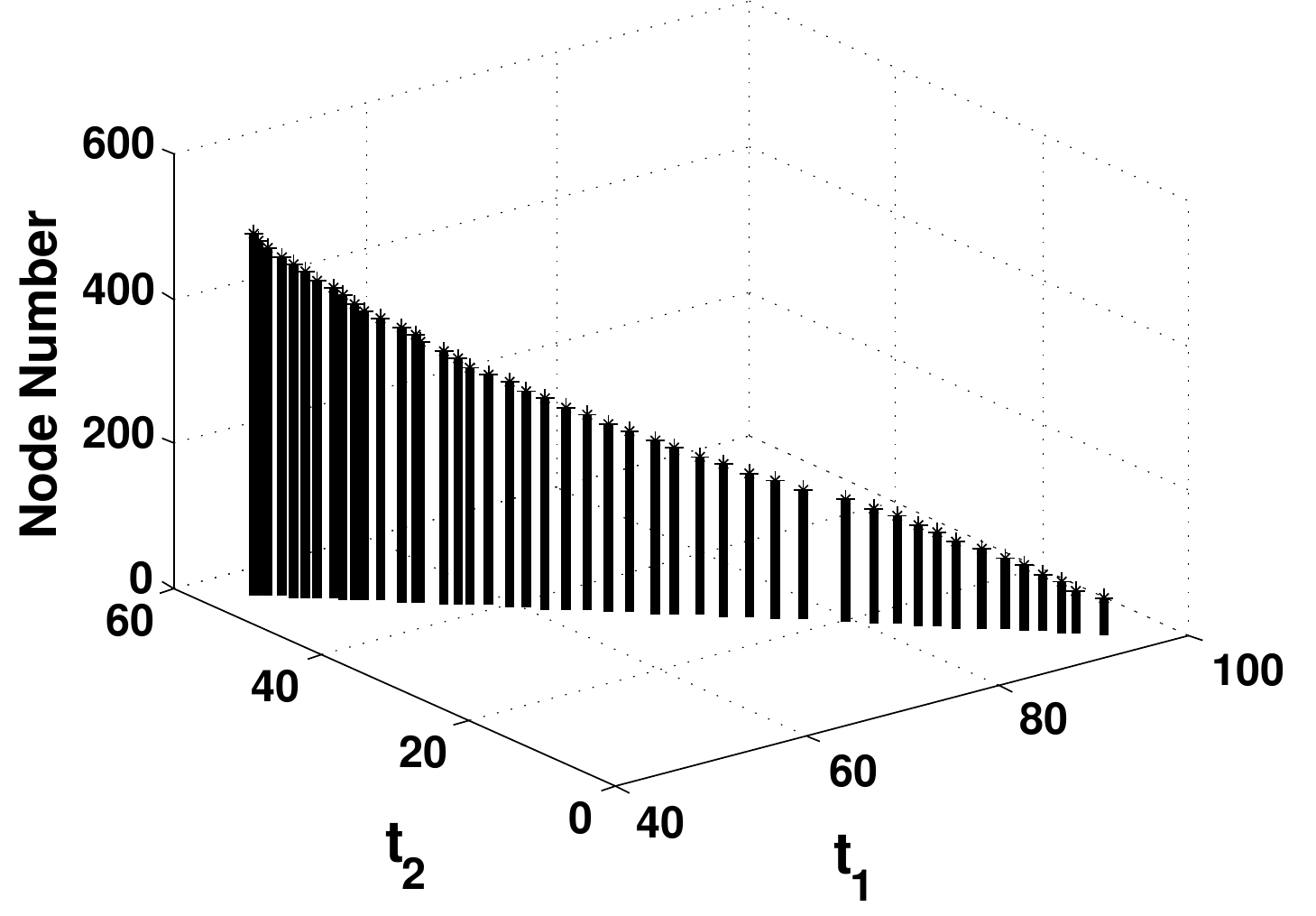}
	\caption{The time percentages ($t_1,t_2$) of algorithm steps for node numbers $50$ to $500$ with an increment 10. $t_1$ is the percentage of label placement time, $t_2$ is the percentage of label-to-node distance calculation time.}
	\label{fig:fig_allnodes_t1t2_nodenum}
\end{figure}

 We have also looked at the relationship between node number and time percentage of the three stages of the labeling algorithm that are given in \rsec{sec:GBLP_algorithm}.  In the experiments we see that the time percentage ($t_3$) of the third part (assignment of labels to the corresponding point features)  is very small (less than 1\%) compared to time percentage of other parts (calculation of potential label positions $t_1$  and obtaining the label-to-node distance matrices $t_2$) and can be neglected. To understand how $t_1$ and $t_2$ change with the node number, for each node set ($n=50$ to $500$ with an increment of ten) we performed 200 trials keeping constant the size of the labels and map and  averaged the results. As seen from the \rfig{fig:fig_allnodes_t1t2_nodenum}, the increment in the number of nodes reduces $t_1$ and increases $t_2$ since the relation between the number of nodes and labels placed on an empty space on the map is not linear. For example a tenfold increase in the node number reduces the area for placing labels only about 15\%. 

 \begin{figure*}[t]
 	\begin{minipage}[t]{0.5\linewidth}
\centering
\includegraphics[width=1\linewidth]{fig_allnodes_t1t2_nodenum}
\caption{The time percentages ($t_1,t_2$) of algorithm steps for node numbers $50$ to $500$ with an increment 10. $t_1$ is the percentage of label placement time, $t_2$ is the percentage of label-to-node distance calculation time.}
\label{fig:fig_allnodes_t1t2_nodenum}
 	\end{minipage}
 	\hspace{0.1cm}
 	\begin{minipage}[t]{0.5\linewidth} 
\centering
\includegraphics[width=0.95\linewidth]{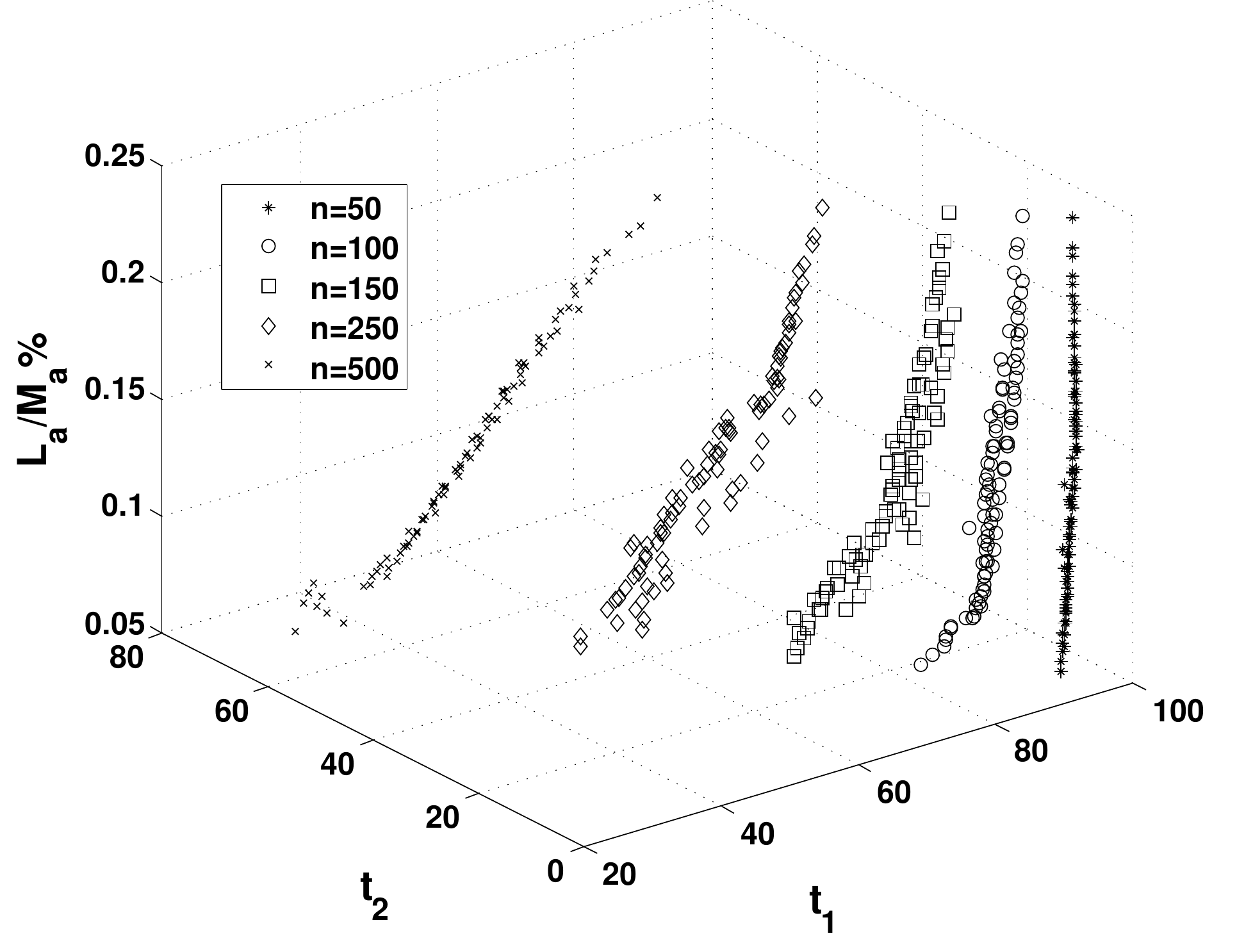}	\caption{The relation between the ratio of label area ($L_a$) to map area ($M_a$)  and time percentages ($t_1,t_2$)  of algorithm steps for node number $50,100,150,250$ and $500$. $t_1$ is the percentage of label placement time, $t_2$ is the percentage of label-to-node distance calculation time.}
\label{fig:fig_allnodes_t1t2_area}
 	\end{minipage}        
 \end{figure*}  
 
\begin{table}[]
	\centering
	\caption{The relation between label size $L_a$ and time percentage of algorithm when the label size is increased five-fold. $t_{1_0}$ and $t_{2_0}$ represent the initial percentages of time.}
	\label{tab:labelsize_timeperc}
	\renewcommand{\arraystretch}{1.75}
	\begin{tabular}{l|c|l|l|l}
		\cline{2-4}
		& \# of node & $\frac{\Delta t_1}{t_{1_0}} \% $  & $\frac{\Delta t_2}{t_{2_0}} \% $  \\ \cline{1-4}
		\multicolumn{1}{|l|}{\multirow{5}{*}{$L_a$x$5$}} & $500$ & $100 \% \uparrow$ &  $40 \% \downarrow$  \\ \cline{2-4}
		\multicolumn{1}{|l|}{}& $250$ & $32 \% \uparrow$ & $49 \% \downarrow $\\ \cline{2-4}
		\multicolumn{1}{|l|}{}& $150$ & $18 \% \uparrow$ & $51 \% \downarrow$\\ \cline{2-4}
		\multicolumn{1}{|l|}{} & $100$ & $10 \% \uparrow$ & $47 \% \downarrow$\\ \cline{2-4}
		\multicolumn{1}{|l|}{}& $50$  & $1 \% \uparrow$ & $33 \% \downarrow$\\ \cline{1-4}
	\end{tabular}
\end{table}
We are also interested in how the total run-time of the algorithm has been affected by node number and label size. \rfig{fig:tavg_h_w_nodes} shows the running time of the algorithm where the height and width of the labels are represented in the x-y direction. We increased $h$ starting from $50$ to $150$, and $w$ starting from $130$ to $200$ with an increment of $10$. We performed $20$ trials for each label size and we repeated this for $n=50,100,150,250$ and $500$. For all node numbers, increasing the label size increases the run time of the algorithm. We also looked the changing in the duration of the algorithm parts. For all node numbers, increasing the label size increases $t_1$, and decreases $t_2$ but the amount of  change is not the same for all sets. For example, when $n=500$ 500, a five-fold increase in label size increases $t_1$ about $100\%$, decreases $t_2$ about $40\%$ but for $n=50$ these values are $1\%$ and $33\%$, respectively. Values for other number of nodes are summarized in \rtab{tab:labelsize_timeperc}. We conclude that the performance of the labeling algorithm is much more sensitive to label size when we increase the number of nodes.



The comparison of our algorithm's performance in terms of
accuracy and computing time are difficult since most of previous
approaches assume that the label touches its corresponding
feature and measured the efficiency of the algorithm by the
number of features labeled in the final solution without considering
time or just consider the algorithm's speed without maximizing
the number of labeled features. Furthermore, they do
not specify the speed and properties of the system that their algorithm
was run on, which is the most important comparison
criteria.

\section{Conclusion}
\label{conc}
The method  can be used to label any feature-based graphs (e.g., data points). We should implement some of the more important labeling rules set forth by
\cite{Imhof.1975} and \cite{Yoeli.1972} in order to appeal to a wide audience. For example, while penalties for overlaps are included in the algorithm, there is no term corresponding to the spacing between labels, which may be important for visual aesthetics. In addition, the labels are horizontally aligned and cannot be tilted. Although this is the case for the majority of labeling problems, there are graphs where a different orientation of the label might be useful (i.e., labeling the different
functional dependence of a time-series graph). Implementing such additional
features and rules can be an important direction for this work.
Currently, the algorithm supports labeling point feature
graphs. 

\section{Acknowledgment}

This research was partially supported by ATOS IT Consulting Customer Service Industrial Trade Co.,Turkey in the scope of AIRC2IS R\&D / Software Development Project.

%
%
\section{References}
  \bibliographystyle{elsarticle-num-names} 
  \bibliography{bib_label}

\end{document}